\documentclass[%
 reprint,
 twocolumn,
nofootinbib,
 amsmath,amssymb,
 aps,
 pra,
floatfix,
]{revtex4-2}

\usepackage{microtype}
\usepackage{algpseudocode}

\usepackage{booktabs}
\usepackage{graphicx,
            caption,
            subcaption,
            ragged2e,
            xcolor,
            xspace,
            textcomp,
            hyperref,
            cleveref,
            scrextend, % for KOMA script functionalities
            }
            
\DeclareCaptionType{algorithm}[ALG.]
\DeclareCaptionLabelSeparator{dot}{. }
\makeatletter
\def\justified{
	\let\\\@normalcr
	\@rightskip\z@skip \rightskip\@rightskip
	\leftskip\z@skip
	\parindent 0em\relax
	\setlength{\parfillskip}{0pt plus 1fil}}
\DeclareCaptionJustification{justified}{\justified}

\hypersetup{colorlinks=true, linkcolor=blue,citecolor=blue,urlcolor=blue}

\usepackage{scalerel}
\usepackage{tikz}
\usetikzlibrary{svg.path}
\definecolor{orcidlogocol}{HTML}{A6CE39}
\tikzset{
  orcidlogo/.pic={
    \fill[orcidlogocol] svg{M256,128c0,70.7-57.3,128-128,128C57.3,256,0,198.7,0,128C0,57.3,57.3,0,128,0C198.7,0,256,57.3,256,128z};
    \fill[white] svg{M86.3,186.2H70.9V79.1h15.4v48.4V186.2z}
                 svg{M108.9,79.1h41.6c39.6,0,57,28.3,57,53.6c0,27.5-21.5,53.6-56.8,53.6h-41.8V79.1z M124.3,172.4h24.5c34.9,0,42.9-26.5,42.9-39.7c0-21.5-13.7-39.7-43.7-39.7h-23.7V172.4z}
                 svg{M88.7,56.8c0,5.5-4.5,10.1-10.1,10.1c-5.6,0-10.1-4.6-10.1-10.1c0-5.6,4.5-10.1,10.1-10.1C84.2,46.7,88.7,51.3,88.7,56.8z};
  }
}
\newcommand\orcidicon[1]{\href{https://orcid.org/#1}{\mbox{\scalerel*{
\begin{tikzpicture}[yscale=-1,transform shape]
\pic{orcidlogo};
\end{tikzpicture}
}{|}}}}

\begin{document}

\author{Maximilian Sohmen~\orcidicon{0000-0002-5043-2413}}
    \thanks{Correspondence should be addressed to\\ \url{maximilian.sohmen@i-med.ac.at}}
\affiliation{Institute of Biomedical Physics, Medical University of Innsbruck, Müllerstraße~44, 6020~Innsbruck, Austria}
\author{Molly A. May~\orcidicon{0000-0001-7164-9569}}
\affiliation{Institute of Biomedical Physics, Medical University of Innsbruck, Müllerstraße~44, 6020~Innsbruck, Austria}
\author{Nicolas Barré~\orcidicon{0000-0002-4460-4151}}
\affiliation{Institute of Biomedical Physics, Medical University of Innsbruck, Müllerstraße~44, 6020~Innsbruck, Austria}
\author{Monika Ritsch-Marte~\orcidicon{0000-0002-5945-546X}}
\affiliation{Institute of Biomedical Physics, Medical University of Innsbruck, Müllerstraße~44, 6020~Innsbruck, Austria}
\author{Alexander Jesacher~\orcidicon{0000-0003-4285-9406}}
\affiliation{Institute of Biomedical Physics, Medical University of Innsbruck, Müllerstraße~44, 6020~Innsbruck, Austria}

\title{Sensorless wavefront correction in two-photon microscopy across different turbidity~scales} 

\begin{abstract}
Adaptive optics (AO) is a powerful tool to increase the imaging depth of multiphoton scanning microscopes. 
For highly scattering tissues, sensorless wavefront correction techniques exhibit robust performance and present a straight-forward implementation of AO. 
However, for many applications such as live-tissue imaging, the speed of aberration correction remains a critical bottleneck. 
Dynamic Adaptive Scattering compensation Holography (DASH) -- a fast-converging sensorless AO technique introduced recently for scatter compensation in nonlinear scanning microscopy -- addresses this issue. 
DASH has been targeted at highly turbid media, but to-date it has remained an open question how it performs for mild turbidity, where limitations imposed by phase-only wavefront shaping are expected to impede its convergence. 
In this work, we study the performance of DASH across different turbidity regimes, in simulation as well as experiments. 
We further provide a direct comparison between DASH and a novel, modified version of the Continuous Sequential Algorithm (CSA) which we call Amplified CSA (a-CSA).
\end{abstract}

\date{\today}

\maketitle

\section{Introduction}

Dynamic wavefront correction is a powerful approach for extending the imaging depth of nonlinear microscopy in scattering tissues such as the mouse brain.
However, to make the benefits of this technology accessible to an even broader range of applications, some remaining limitations need to be overcome.
A particular problem is posed by the speed at which an aberration-compensating pattern can be retrieved, as imaging into live specimens requires to outpace the decorrelation time imposed by the constantly changing tissue. 
\newline
To this end, we have recently introduced Dynamic Adaptive Scattering compensation Holography (DASH), a fast-converging indirect wavefront sensing algorithm for scatter correction in nonlinear scanning microscopy~\cite{may2021fast}. 
DASH employs a programmable phase-diffractive element to shape two beams simultaneously: 
a test beam, whose wavefront is varied in each step to explore possible signal improvements, and a corrected beam, whose wavefront is continuously improved using the information gained from interference with the test beam. 
DASH employs phase-only wavefront shaping, which bears the advantage of power efficiency. 
On the other hand, discarding amplitude information introduces errors in the resulting wavefronts.
In this work we investigate the impact of these errors on the correction performance of DASH for different regimes of turbidity. 
Furthermore, we compare DASH to an alternative approach, which does not suffer from the ``phase-only'' restriction: a modified version of the Continuous Sequential Algorithm (CSA)~\cite{vellekoop2008phase}, which follows a pixel-by-pixel testing approach. 
Our modification to CSA consists in amplifying the intensity of the tested pixel relative to all other pixels, which increases signal contrast. 
This amplification allows the application of CSA in situations with low signal-to-noise ratio (SNR), as often encountered in practice. 
We refer to our modified CSA algorithm as \emph{Amplified} Continuous Sequential Algorithm, or a-CSA.

This work is structured as follows: In Section~\ref{sec:turbulence} we discuss how turbidity can be quantified, how a scattering medium of specific turbidity can be modelled numerically,
and how in a two-photon imaging experiment a scattering volume of `tunable' mean free path can be emulated by a phase-mask displayed on a spatial light modulator (SLM).
In Section~\ref{sec:methods}, the principles behind DASH and a-CSA are explained. 
In Section~\ref{sec:comparison}, we detail the implementation of a-CSA and DASH in our numerical simulations (Section~\ref{sec:simu}), our experiment (Sections~\ref{sec:experiment}), and present a comparison of algorithm performance for two-photon excited fluorescence (TPEF) imaging of a homogeneous `dye-slide' sample in different regimes of turbidity (Section~\ref{sec:systcomp}) as well as mouse brain (Section~\ref{sec:brain}).
In Section~\ref{sec:lowSNR}, implications of a low SNR on algorithm performance are discussed. 

\section{Quantifying turbidity}
\label{sec:turbulence}

It is essential for the present work to define what we mean when speaking of ``low'' or ``high'' turbidity. 
The scattering properties of materials and tissues are often quantified using the scattering mean free path $l_s $, i.e., the expectation value of a photon's free travelling path before it is scattered.
This is mirrored in the Beer-Lambert law, $|U_0(L)|^2 = |U_0(0)|^2~\mathrm{e}^{-L/l_s}$, where $|U_0(L)|^2$ represents the intensity of the \emph{unscattered} (`ballistic') light after travelling (under free-space propagation) to distance $L$, and $|U_0(0)|^2$ the incident light intensity.
The transport mean free path $l_t$ takes scattering anisotropy into account: $l_t = l_s/( 1-g)$, where $g = \langle\cos\theta\rangle$ is the expectation value of the cosine of the scattering angle $\theta$. For instance, in a material which predominantly scatters into the forward direction (causing small scattering angles), $l_t$ is much larger than $l_s$. 
Conversely, in an isotropic scatterer $l_t = l_s$.
{\color{black}Typical values of $l_s$ for brain tissue range between a few tens to hundreds of micrometers~\cite{oheim2001two, Chaigneau:11,jacques2013optical}.}

Our goal is to model the effect of a (in general \emph{three-dimensional}, 3D) scattering medium on a light field propagating in positive $z$-direction by a \emph{two-dimensional} (2D) phase mask, located at axial position $z=z_\text{scat}$, with transmission function $\exp\left(\mathrm i\Phi(\boldsymbol\rho)\right)$.
Here, $\Phi(\boldsymbol\rho)$ denotes the scattering-related phase shifts experienced by a field point at the 2D lateral coordinate $\boldsymbol\rho$. 
The field after the phase mask is denoted by $U_\text{scat}(\boldsymbol{\rho})$. 
Note that this is the full field, not just a `scattered field' amplitude. 
Assuming that the phase mask is suitably chosen to describe a medium with predominantly forward scattering and without absorption, we choose the normalisation
\begin{equation}
 \int |U_\text{scat}(\boldsymbol{\rho})|^2 \,\mathrm d\boldsymbol{\rho} = 1 = \int |U_0(\boldsymbol{\rho})|^2 \,\mathrm d\boldsymbol{\rho}.   
\end{equation}
\newline
The ballistic contribution at depth $L$ inside the medium -- emulated by the phase mask on the SLM -- can be calculated using the overlap integral (OI)
\begin{align}
    \text{OI}\left[U_\text{scat}, U_0\right] 
    &= \left|\int U_\text{scat}(\boldsymbol{\rho})\, U^*_0(\boldsymbol{\rho})\,\mathrm d\boldsymbol{\rho}\right|^2 \\
    &= \left|\int \left|U_0(\boldsymbol{\rho})\right|^2 \mathrm e^{\mathrm i\Phi(\boldsymbol{\rho})}\,\mathrm d\boldsymbol{\rho}\right|^2,
    \label{eq:1}
\end{align}
i.e., the `projection' of the field with imprinted phase mask onto the unscattered (incident) field. This equality~(\ref{eq:1}) is most intuitive if the integral is evaluated in the plane of the 2D scattering mask, but for freely propagated fields the OI in fact stays constant in all transverse planes at $z \geq z_\text{scat}$.
Using the Lambert-Beer law, the OI can also be written as
\begin{equation}
    \text{OI}[U_0(L), U_0(0)] 
    = \mathrm e^{-L/l_s}.
    \label{eq:OI2}
\end{equation}
$l_s$ appears here, since every single scattering event reduces the ballistic contribution.  
Note that this relation~(\ref{eq:OI2}) implicitly assumes that cases of successive scattering events which exactly compensate each other (thus, re-populating the forward-directed incident field, i.e., contributing to the OI and -- erroneously -- to the estimated ballistic part) are statistically unlikely and can be ignored.
\newline
Combining Eqs.~1--2 we can quantify a computed phase mask in terms of the corresponding ``thickness'' expressed in units of the scattering mean free path $l_s$:\footnote{We note that the relation~(\ref{eq:3}) is consistent with the considerations made in Ref.~\cite{wang2011scattering} (see Eq.~4 therein), which lead to the derivation of the scattering-phase theorem.} 
\begin{equation}
    L/l_s 
    = -\ln(\text{OI})
    = -\ln\left(\left|\int \left|U_0(\boldsymbol{\rho})\right|^2 \mathrm e^{\mathrm i\Phi(\boldsymbol\rho)}~\mathrm d\boldsymbol\rho\right|^2\right)
    \label{eq:3}
\end{equation}
For the case of dominant forward scattering and negligible absorption, this relation allows us to compute a 2D phase mask $\Phi(\boldsymbol\rho)$ that leads to a speckle pattern in the object plane which is in many ways similar to that of a voluminous 3D scatter medium of the same scattering mean free path $l_s$. 
In the experiments described later in this work, we will exploit this fact to simulate different regimes of turbidity by displaying computed 2D scatter masks of specific $l_s$ on an SLM.
Of course the equivalence between a 3D and a 2D scatterer -- even if they exhibit the same $l_s$ -- does not encompass all physical properties; for instance, the isoplanatic patch (i.e., the `corrected field of view') obtained through an AO wavefront correction will be smaller for a 3D than for a 2D scatterer.
However, concerning the aspects studied in this work (e.g., the algorithm convergence at a single field point), a 3D and a 2D scatterer of same $l_s$ can be regarded as equivalent.

\noindent
We denote the RMS value of a scattering phase mask by $a_\text{scat}$ (see Algorithm~4, Supplementary Material). 
If the phase values of the mask are normal-distributed or, for any distribution, if $a_\text{scat}$ is sufficiently small~\cite{wang2011scattering}, the relation between the scatterer thickness and $a_\text{scat}$ is simply $L/l_s = \sqrt{a_\text{scat}}$.

\section{Approaches for sensorless AO in nonlinear microscopy}\label{sec:methods}

Most indirect (or sensorless) AO schemes construct an aberration compensation phase mask from measurements of the TPEF signal for many different test patterns or ``modes'' $M_n$ displayed on an SLM, $n \in \{1,\ldots, N\}$ denoting the mode index. 
These test patterns can be pre-defined, for instance as a set of Zernike~\cite{Booth:02} or Hadamard modes~\cite{conkey2012high, Blochet:17}, or directly derived from the signal of previous test patterns, such as in genetic algorithms~\cite{conkey2012genetic}.
The phase patterns are commonly imaged into the back focal plane (BFP) of the objective lens, where -- by the Fourier transform property of the lens -- the test patterns have a homogeneous effect on the point-spread function (PSF) over the entire focal plane.
However, there exist also sample-conjugated schemes, where the SLM is imaged directly into the sample space~\cite{paudel2015axial,Tao:17,papadopoulos2020dynamic, may2021simultaneous}.
\begin{figure*}[tbh]
    \centering
    \includegraphics[width = 0.9\textwidth]{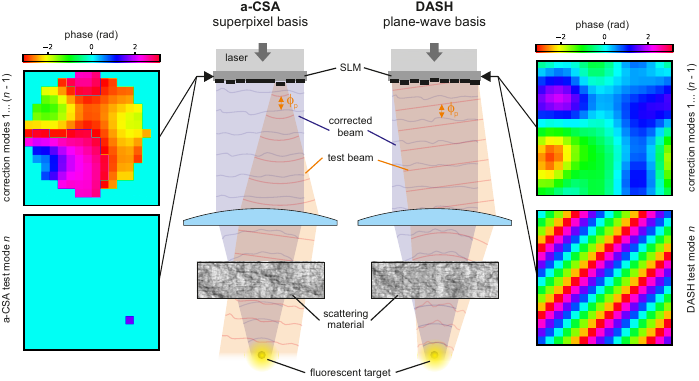}
    \caption{\textbf{Sensorless AO methods for nonlinear microscopy.} a-CSA uses a superpixel basis, DASH a plane-wave basis of test modes on a fluorescent target. Respective examples of a test mode (bottom) and the compensation pattern at the corresponding step are shown on the left and right. 
    For a-CSA the order of tested modes is by increasing distance to the pupil center, for DASH the order is by increasing spatial frequency.
    During phase-stepping, the relative phase $\phi_p$ of the test mode with respect to the compensation pattern is varied (see main text).}
    \label{fig:concepts}
\end{figure*}

A particular case is the ``traditional adaptive optics'' regime, where the given phase aberration $\Phi(\boldsymbol\rho)$ can be approximated by a weighted sum of the available test modes, $\Phi(\boldsymbol\rho) = \sum_{n = 1}^N a_n\,M_n(\boldsymbol\rho)$, and each $a_n$ has a rather small phase magnitude (up to about 1~rad RMS over the pupil). 
Here, $\boldsymbol\rho$ is the 2D coordinate vector in the BFP. 
In this case, the TPEF signal $S$ generated in the focus has a smooth dependence on the mode magnitudes $a_n$, allowing to approximate it by a simple function such as a multi-dimensional Lorentzian or parabola, e.g., $S \propto 1 - \sum_n\sum_m \alpha_{nm}\,a_n\,a_m$. 
In many cases, the cross-talk matrix $\alpha$ can be diagonalized by choice of an adequate mode basis~\cite{kubby2013adaptive}. 
Then, $N+1$ measurements can be sufficient to characterize the phase aberration~\cite{Booth:06}, although usually $2N+1$ or even $3N$ measurements are taken. 
It is important to note that this particular case does not necessarily coincide with low turbidity (i.e., a small value of $L/l_s$), since a large number of modes, even if their individual magnitudes are small, can still sum up to a large total aberration.  
\newline
Outside the traditional regime, in what is often called the ``scattering'' regime, it is usually required to take many more measurements to find a suitable corrective phase mask. 
Typically, in the scattering regime the complexity of the aberration is beyond the capabilities of the correction device, meaning that the number of scattering modes is larger than the number of correctable ones. 
Additional reasons which can hinder full aberration compensation include an SNR too small to measure the contributions of less significant modes, or simply lack of time to measure all of the (many) scattering modes. 
Nevertheless, it has been shown that even correcting only a limited number of scattering modes can suffice to restore a diffraction-limited focus, which in this context can be understood as an intensity-enhanced speckle~\cite{Vellekoop:07}.
\newline
There exist several approaches for sensorless AO which can operate in the scattering regime~\cite{vellekoop2008phase, tang2012superpenetration, conkey2012genetic, papadopoulos2017scattering, rodriguez2021adaptive, may2021fast, blochet2021fast}.
Two of them, which we have identified as fast converging in numerical simulations, are studied in this work and outlined in Fig.~\ref{fig:concepts}.

The first approach, known as continuous sequential algorithm (CSA)~\cite{vellekoop2008phase}, operates on a single-pixel basis: sequentially, each pixel's phase is adjusted to maximize the signal. 
Although a basis of single pixels appears advantageous due to its intrinsic orthogonality, 
the approach is usually impractical because the interference contrast from single-pixel phase modulation is typically buried in noise. 
This inspired alternative approaches, such as hybrid methods using larger ``superpixels'', each of them featuring an internal spatial phase pattern~\cite{rodriguez2021adaptive, hu2020universal}.
However, we point out that the SNR problem of CSA can also be remedied by simply amplifying the intensity in the test superpixel, as this increases the interference contrast. 
This variant, which we call a-CSA (``amplified CSA''), requires the laser power on each superpixel to be controllable.
Then, as we show in Section~\ref{sec:experiment}, a-CSA can work in experiments where only a comparably small number of modes needs to be corrected.

\begin{figure}
    \centering
    \includegraphics[width = \columnwidth]{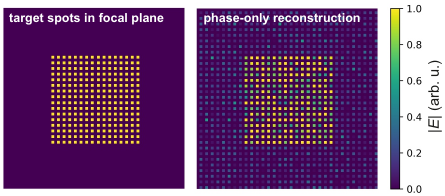}
    \caption{\textbf{Artefacts related to phase-only light shaping.} 
    Field modulus of a target pattern (left) and the simulated pattern reconstructed by phase-only wavefront shaping in the Fourier plane (right), illustrating introduced artefacts.}
    \label{fig:artefacts}
\end{figure}

The mode basis of the second approach, DASH, is a set of plane waves which are, in a sense, Fourier-related to the single-pixel basis. 
Importantly, for this approach the SNR is uncritical, as the power fraction contained in each plane wave can be adapted directly via the SLM hologram. 
This makes this method more practical than CSA. 
However, to shape test beam and corrected beam without introducing artefacts, the SLM would be required to manipulate both, the spatial phase and amplitude distribution of the incident laser beam, just as for a-CSA.
In DASH, the amplitude part is disregarded and only the phase part is included in the hologram, ensuring high power efficiency but inevitably leading to errors in the generated wavefronts and undesired diffraction orders. 
An example of the effects of phase-only shaping is given in Fig.~\ref{fig:artefacts}. 
The left image shows a target arrangement of 256 spots with random phases we wish to create in the focal plane using an SLM in the Fourier plane. 
Discarding the amplitude part of the synthetic hologram (a superposition of the corresponding 256 plane waves) leads to the image on the right. 
The standard deviation of the phases at the target sites is about 0.25\,rad, corresponding to about 4\% of the wavelength, and the relative amplitude error at the target sites is about 20\%.
Additionally, we see that weak ``ghost spots'' appear outside the target square. 

In line with our earlier works~\cite{may2021fast, may2021simultaneous}, we will in the following for both a-CSA and DASH denote by $f$ the fraction of the total pupil intensity which is contained in the test mode.

\section{Comparison of algorithm performance}
\label{sec:comparison}

In this section we compare the performance of DASH and a-CSA over different scales of turbidity in numerical simulations and experiment.

\subsection{General recipe}

The procedure for our systematic comparison between DASH and a-CSA is as follows.
\begin{enumerate}
    \item Both algorithms are initialised with an empty correction mask.
    \item We calculate a scattering phase mask that emulates a certain degree of turbidity and is kept constant throughout each algorithm run. 
    In the simulations, this scattering mask is simply added to the BFP-conjugate wavefront during image propagation; in the experiment, it is displayed on our SLM.
    \item\label{step:test} 
    We imprint a test mode on the excitation beam and carry out a phase-stepping procedure, i.e., measure the TPEF signal generated in the object plane for $P$ different global phase offsets $\phi_p = 2\pi\,p/P$ (with $p = 1\ldots P$) applied to the test mode.\footnote{In the experiment we typically set $P = 5$; in the simulations, especially for the noise-free case, we find that $P$ can be reduced to $3$ for faster computation without loss of performance.}
    From the phase-stepping procedure, we extract the parameters for which the test mode maximizes the TPEF signal.
    \item We directly update our correction mask by including the tested mode with the retrieved optimal parameters. In the simulations, this correction mask is added to the wavefront incident on the scatterer; in the dye-slide experiment, it is displayed on our SLM, on top of the scattering mask. 
    \item We now go back to step~\ref{step:test}, test the next mode, update the correction mask, and so forth, until the correction mask contains the full number of test modes with their optimal contributions.
    \item Once the full set of modes has been tested, we can start again from the first test mode for another correction run. 
    We typically run 3 full iterations, which has shown to ensure algorithm convergence in most cases.
\end{enumerate}
Detailed algorithm descriptions are provided in the Supplementary Material.\footnote{This also includes a minor modification to the DASH algorithm compared to Ref.~\cite{may2021fast}.}

\subsection{Numerical model}
\label{sec:simu}

In our numerical model, we consider aberrations defined on a square grid of 64${\times}$64 pixels accounting for $N^2_\text{scat} = {}$4096 scattering modes.
The aberration phase masks are calculated by adding cosines of all spatial frequencies supported by the grid size with uniformly distributed random phases.
The amplitudes of the cosines follow a Gaussian weighting with standard deviation $\sigma$, such that cosines with higher spatial frequencies are increasingly attenuated, as expected for most scattering scenarios found in nature. 
By varying $\sigma$ we can hence tune the spatial frequency content of the model scatterer. 
For details on our specific implementation of calculating the scatter mask, the reader is referred to our Supplementary Material (Algorithm~4).

The simulated SLM for (square) correction patterns features 32${\times}$32 pixels. 
Both the scattering mask as well as the simulated SLM are located in the BFP, in Fourier relation to the object plane, and have a side length measuring $2
k_0\,\text{NA}$, where $k_0$ is the vacuum wavevector of light and $\text{NA} = {}$0.7 the numerical aperture.
Each set of modes is tested repeatedly for 3~iterations. 
In our experiments, the SLM illumination does not feature an ideal, flat-top intensity profile, but exhibits a (weak) Gaussian shape of $\mathrm{e}^{-1/2}$~distance approximately equal to the pupil radius.
To reflect the experimental situation, also in our simulations we assume a light intensity distribution across the BFP which is symmetrically Gaussian and of corresponding width.
\newline
The fluorescent sample is assumed to be a homogeneous, absorption-free fluorescent layer of $d_s = 10$\,µm thickness.\footnote{NB that a thin 3D fluorescent sample layer as assumed here leads to smaller signal enhancements and slower algorithm convergence than an (infinitely thin) 2D layer. 
In the limit of a homogeneous 3D sample volume with thickness much larger than the Rayleigh range of the focused beam, the second-order nonlinearity of TPEF imaging is insufficient for providing any signal enhancement at all~\cite{katz2014non}.}
We model this sample volume by grid points on $N_s$ distinct 2D planes with axial interspaces of $d_s/(N_s - 1)$, and the nominal focus plane located at its center. 
Normally, we set $N_s = 6$ layers (interspaced by 2\,µm), since a larger number of planes does not notably improve the accuracy of results and we have checked that the exact choice of $N_s$ is uncritical for our conclusions.
To simulate the PMT counts, we simply propagate the light field into the $N_s$ planes, calculate the squared intensity on each grid point, and take the sum over all points.

In Section~\ref{sec:systcomp}, we show simulation results at high-SNR conditions, for which photon shot noise can be neglected.
For scenarios with low SNR, as discussed in Section~\ref{sec:lowSNR}, we simulate shot noise by varying the summed PMT counts according to Poisson statistics.

\subsection{Experimental procedure}
\label{sec:experiment}

\begin{figure*}[tbh]
    {\centering
    \includegraphics[width = 0.75\textwidth]{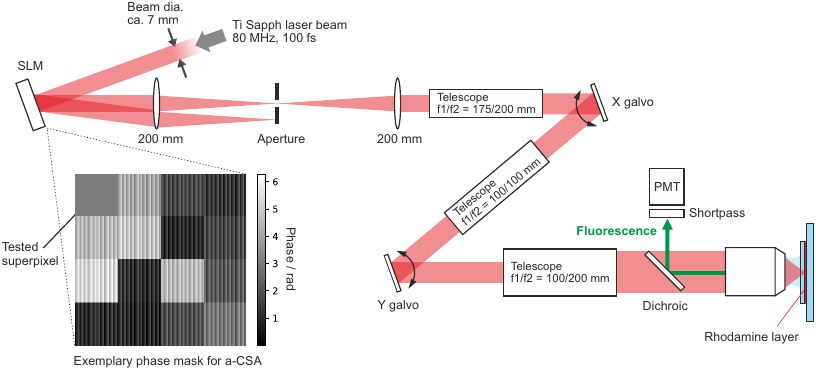}}
    \caption{\textbf{Experimental setup.} 
    SLM = spatial light modulator; PMT = photomultiplier tube.
    Inset: Exemplary test phase mask for a-CSA. Every superpixel except the one under test is superposed by a shallow blazed grating to diffract the desired fraction of light intensity off the optical axis where it is blocked by an iris aperture. This enables amplitude tunability while using a standard, phase-only SLM.}
    \label{fig:setup}
\end{figure*}

We compare DASH and a-CSA experimentally in a home-built TPEF microscope featuring a phase-only SLM (Meadowlark HSP1920-500-1200-HSP8, 1920${\times}$1152 pixels of side length 9.2\,µm) located in a BFP-conjugate plane of the excitation path. 
This SLM serves two purposes in parallel:
First, by displaying a `scattering' phase mask of defined scattering mean free path $l_s$ (see Section~\ref{sec:turbulence}) it allows to \emph{emulate} the effect of a scattering medium in the light path. 
Second, by running our sensorless wavefront correction schemes and displaying the retrieved phase compensation patterns \emph{on top} of the given scattering mask, we can test and compare the algorithm performances.
\newline
A sketch of the setup is shown in Fig.~\ref{fig:setup}.
For excitation, we use a mode-locked Ti:sapphire laser (MaiTai DeepSee, Spectra Physics) with emission maximum set to 800\,nm wavelength. 
The epi-TPEF is collected by a water-dipping objective (Olympus XLUMPLFLN20XW, $\text{NA}=1$) and directed towards a photomultiplier tube (PMT, Hamamatsu H10769A-40) using a dichroic beam splitter (AHF Analysentechnik, HC665LP) and an additional emission filter in front of the PMT (AHF, ET680SP-2P8).
In our measurements, to operate with a more homogeneous pupil illumination, we artificially reduce the NA to about 0.7 using an aperture in a pupil-conjugate plane (not shown in Fig.~\ref{fig:setup}). 

For our systematic comparison, the sample consists of a thin layer of rhodamine solution sandwiched between a glass slide and a coverslip of 1\,mm and 170\,µm thickness, respectively. 
Before each algorithm run, we perform an initial precorrection run to compensate for optical-system wavefront distortions such as the spherical aberrations introduced by the coverslip.
Starting from this precorrection, the wavefront correction algorithms afterwards only have to correct for the artificial scatterer displayed on our SLM, ensuring compatibility with the numerically simulated correction runs.

For DASH, we typically set the intensity in the test mode to 30\% of the total intensity (i.e., $f = 0.3$, see Supplementary Material).
For a-CSA, we amplify the relative power in the test-superpixel by superposing \emph{all other} a-CSA-superpixels (consisting of many physical SLM pixels) with a blazed grating of defined modulation depth~\cite{bagnoud2004independent}.
Thus, for all except the test-superpixel, power is diffracted away from the optical axis and cut by an aperture as indicated in Fig.~\ref{fig:setup}.
It is important that the mean phase of the blazed grating is nought to prevent an effect on the zero-order beam, which carries the phase mask.\footnote{Note that by encoding our phase masks in the zero order and dumping power in higher orders we avoid dispersion of the femtosecond laser pulses.}
Our blazed gratings have a period of 4~SLM-pixels and a modulation depth of approximately $\pi$\,rad, resulting in $(1-\beta) \sim 50$\% of the laser power being dumped (measured value) and a power fraction $f \approx 1/(\beta N^2)$ in the case of $N{\times}N$ superpixels.
Naturally, dumping power on the iris also decreases the total TPEF signal, wherefore we have to compensate by increasing the total laser power. 
Since in practice the available power is necessarily finite, this approach for amplitude modulation is only applicable in regimes of turbidity which do not require too many (i.e., too small) a-CSA-superpixels.

For compatibility, we conduct our experiments with the same number of test modes as in the simulations for the different regimes, i.e.~16, 64, and 256 modes for low, medium, and high turbidity, respectively.
Our SLM holograms measure 560$\times$560 SLM-pixels and entirely fill the reduced objective pupil. 
For a-CSA this means that we form 4$\times$4 square superpixels (each 140$\times$140 SLM-pixels) for the low turbidity, 8$\times$8 superpixels (each 70$\times$70 SLM-pixels) for the medium turbidity, and 16$\times$16 superpixels (each 35$\times$35 SLM-pixels) for the high turbidity scenario. 
\newline
Analogous to the simulation, DASH and a-CSA are executed for three full iterations.
The results are summarized in Fig.~\ref{simu_A}.

\subsection{Results systematic comparison}
\label{sec:systcomp}

For our systematic comparison, we study three different scenarios (A, B, C) concerned with increasing levels of turbidity. 

\begin{figure*}
    \centering
    \includegraphics[width = 0.9\textwidth]{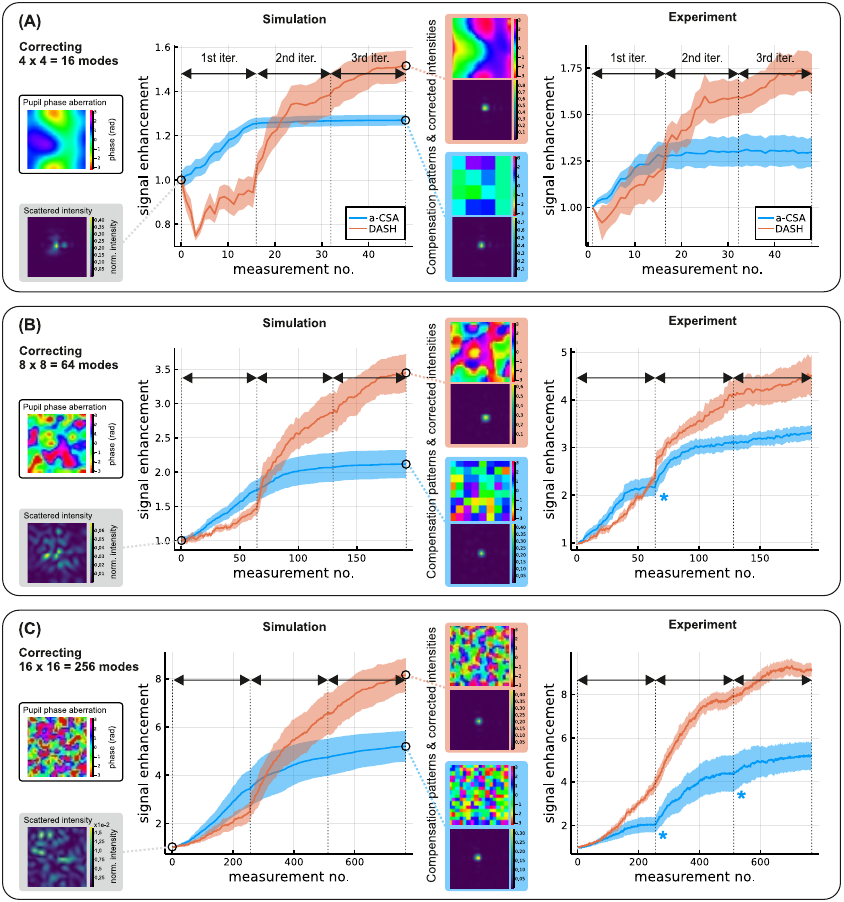}
    \caption{\textbf{Scatter correction in different turbidity regimes.} 
    The three scenarios \textbf{A}--\textbf{C} correspond to an increasing degree of scattering with (\textbf{A}) $L/l_s = 1, \sigma = 1$, (\textbf{B}) $L/l_s = 3, \sigma = 3$, and (\textbf{C}) $L/l_s = 5, \sigma = 5$, respectively.
    The left column shows examples of the phase aberration applied in the objective pupil and the resulting TPEF intensity distribution in the focal plane. 
    The plots in the center and right column show how the TPEF signal improves during correction in simulation (center) and experiment (right).
    Here, solid curves represents the mean value over 5 repeated runs in the simulation or experiment, each initialised with a different random scatterer, and the ribbons represent the respective standard error of the mean. 
    Kinks in the experimental a-CSA curves (blue asterisks) are caused by the circular, underfilled pupil (see main text).
    Additional information, such as retrieved SLM compensation masks and resulting focal spots, is provided in the Supplementary Material.}
    \label{simu_A}
\end{figure*}

In Scenario~A we study low turbidity, with an effective scatterer thickness of a single scattering mean free path, $L/l_s = 1$, and a spatial frequency distribution of the scatterer chosen accordingly narrow, $\sigma = 1$.
An example of such a scatter mask is shown on the top left of Fig.~\ref{simu_A}\,(A).
In the focal plane, such a mild scatterer typically leads to an intensity distribution which is still spatially confined and only moderately deviates from the aberration-free focus, as shown in Fig.~\ref{simu_A}\,(A), bottom left.
Here, the intensity scale is normalized to the peak intensity of the aberration-free focus, such that the maximum value (typically around 0.4 for Scenario~A) equals the respective Strehl ratio.
In this low-turbidity case it is expected that only a small number of modes is needed for adequate compensation, wherefore we correct $4 {\times} 4 = 16$ modes.
The plots in the center and right column of Fig.~\ref{simu_A}\,(A) show the TPEF signal enhancement simulated numerically (center) and measured experimentally (right), respectively, for a-CSA (blue) and DASH (orange). 
Specifically, we plot the TPEF signal measured when the established compensation pattern is applied at the respective measurement number.\footnote{Note that these curves are often similar, but in general not identical to the signal collected during an algorithm run, as the latter results from an interference between a partly compensated and a test beam.}
\newline
In the numerical simulations [Fig.~\ref{simu_A}\,(A), center], we observe that for a-CSA the signal initially increases fast, but then flattens off from the second iteration onwards.  
DASH, in contrast, exhibits a less monotonic signal evolution than a-CSA, first increasing more slowly, but ultimately surpassing the final signal level of a-CSA.
We attribute the signal flattening and lesser total performance of a-CSA to the fact that its compensation patterns are inherently displayed at the native resolution of the test-mode basis (i.e., 4$\times$4 superpixels in Scenario~A) and therfore do not exploit the full resolution supported by the SLM (32$\times$32 pixels in the simulation). 
This coarser spatial discretization leads to increased diffraction losses compared to DASH (which always exploits the full SLM resolution, regardless of the number of tested modes) during the course of the optimisation.
Additionally, we observe that for DASH the steepest changes in signal usually occur at the beginning of each iteration. 
This is due to the fact that we intentionally sort our test modes in ascending order with regard to the angle between their propagation direction and the optical axis.
Since small scattering angles are typically more dominant in fluorescence imaging settings, this tends to speed up the algorithm convergence.
Both methods achieve comparable final Strehl ratios around 0.75 and 0.85 for a-CSA and DASH, respectively (see Supplemental Material).
\newline
Our experimental measurements [Fig.~\ref{simu_A}\,(A), right] show the same general trends.
When comparing absolute values of signal enhancement, it is important to keep in mind the critical dependence on the sample thickness: in our simulations, e.g., we observe for increasing the thickness as 0~$\rightarrow$~4\,µm~${\rightarrow}$~10\,µm (at constant fluorophore density) a tendency of decreasing final enhancements of 
2.14~$\rightarrow$~1.32~$\rightarrow$~1.29 in case of a-CSA and 2.9~$\rightarrow$~1.8~$\rightarrow$~1.5 in case of DASH.
Given that in our experiments the sample layer thickness is controllable and measurable only with limited accuracy, it seems fair to claim good agreement between the observed values in simulation and experiment.

In Scenario~B, we assume medium turbidity with $L/l_s = 3$ and an intermediate contribution of modes of higher spatial frequency, $\sigma = 3$. 
We increase the number of correctable modes to 64 to account for this greater spatial frequency content. 
Before correction, the typical Strehl ratios in this scenario are on the order of 5\% [Fig.~\ref{simu_A}\,(B), bottom left]. 
\newline
Our numerical simulations [Fig.~\ref{simu_A}\,(B), center] indicate that again, a-CSA initially improves faster than DASH, but levels off at a lower final signal.
The total signal enhancement achievable by both correction algorithms is higher than for the low-turbidity case; the final Strehl ratios are around 0.4 for a-CSA and 0.6 for DASH.
We suspect that the initial delay of DASH in comparison to a-CSA is related to the appearance of higher diffraction orders (``ghost foci'' as shown in Fig.~\ref{fig:artefacts}) stemming from the phase-only field shaping of DASH.
This is one of several reasons which lead us to believe that the use of a complex field-shaping technique might bear great potential for improvement of algorithm performance in the future. 
\newline
In our experimental measurements [Fig.~\ref{simu_A}\,(B), right], we observe similar behaviour. 
Initially, a-CSA improves faster than DASH, but ultimately DASH delivers higher signal enhancement. 
For a-CSA, we observe a kink in the signal enhancement at the transition between two iterations (marked by the blue asterisk in Fig.~\ref{simu_A}), which is not present in the simulations. 
We attribute this kink to the circular pupil in our experiment which cuts power from superpixels located near the corners of the square SLM pattern. 
Since we step through the a-CSA superpixels in order of their distance to the pupil center, this effect is most dominant towards the end of each iteration.

In Scenario~C, we assume high turbidity, with $L/l_s = 5$ and $\sigma = 5$, where without correction typical Strehl ratios are on the order of 1\%.
In this scenario, we correct 256 modes.
\newline
In the numerical simulations [Fig.~\ref{simu_A}\,(C), center], we observe that the initial speed advantage of a-CSA compared to DASH decreases, but the final enhancements achievable with both algorithms are even higher than in Scenario~B.
Comparing between the two algorithms, our simulations again deliver a better performance for DASH than for a-CSA.
Final Strehl ratios are around 0.35 for a-CSA and 0.45 for DASH.
\newline
These trends are well supported by our experimental data [Fig.~\ref{simu_A}\,(C), right]. 

Graphical animations of our simulated correction runs are provided as GIF files (see Supplementary Material). 
These animations show the phase patterns displayed on the SLM during the correction algorithm as well as the evolution of the focal plane intensity distribution.

\subsection{Experimental comparison for a biological sample}
\label{sec:brain}

Both DASH and a-CSA can offer striking quality improvements for imaging through turbid media, such as in microscopy of layers deep inside tissue. 
To demonstrate this, we use our TPEF scanning microscope to image microglia expressing green fluorescent protein (\hbox{CX\textsubscript{3}Cr-1\textsuperscript{GFP}}, cf.~Ref.~\cite{may2021fast}), located 200\,µm deep 
inside the corpus striatum of a mouse brain slice fixed via perfusion. 
For this, we adjust our wavelength to the excitation maximum at 900\,nm, retrieve a precorrection for optical system aberrations by focussing on microglia directly below the coverslip, and subsequently move the objective focus mechanically 200\,µm down into the brain tissue.

\begin{figure*}[tbh]
    \centering
    \includegraphics[width = 0.75\textwidth]{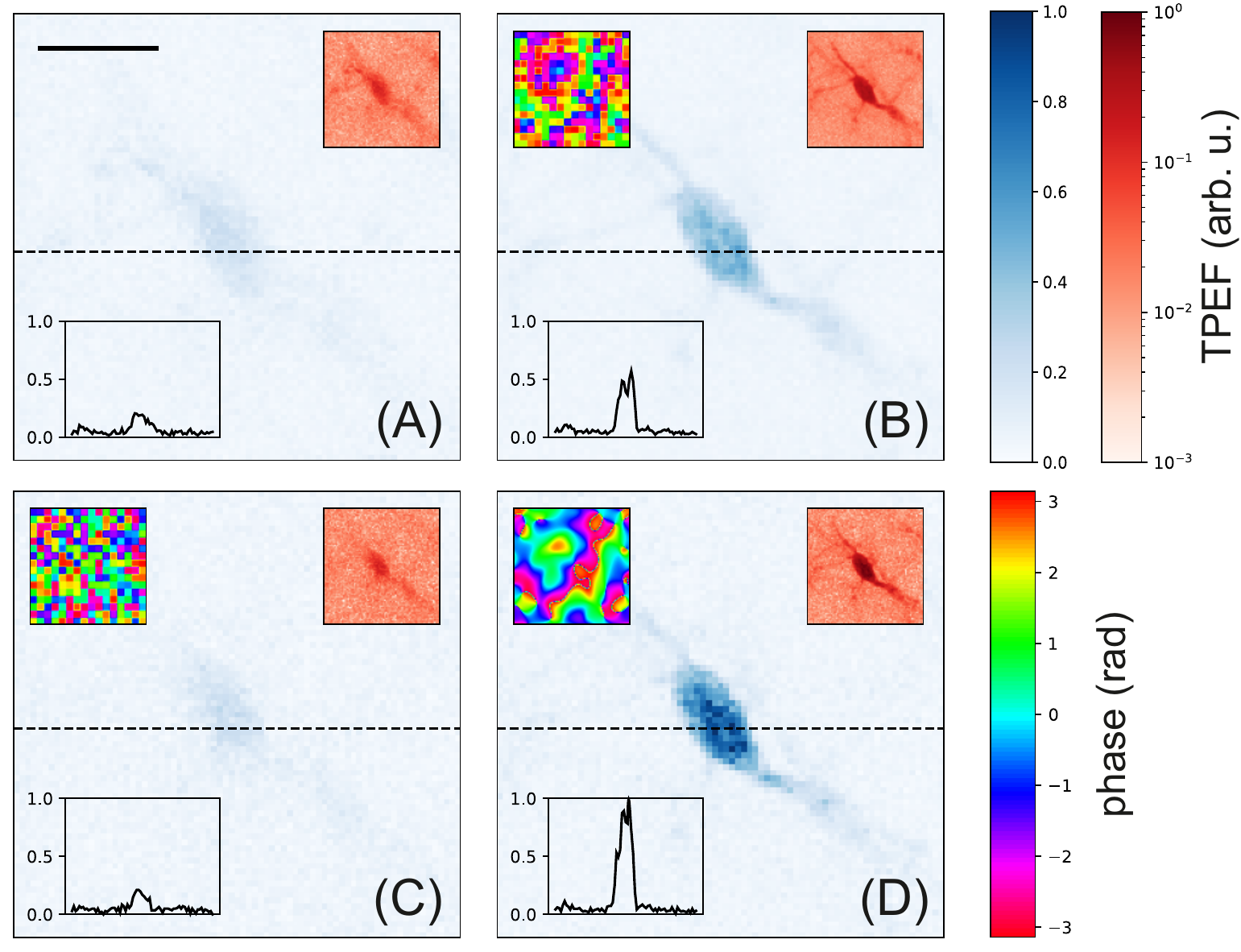}
    \caption{\textbf{Image enhancement for a biological sample.} 
    Murine microglia expressing green fluorescent protein, $200$\,µm deep inside fixed brain tissue.
    Image \textbf{(A)} is recorded using only a precorrection for optical system aberrations. 
    \textbf{(B)}--\textbf{(D)} show the same cell after precorrection plus a subsequent correction (3 iterations of 256 modes) for scattering inside the tissue:
    \textbf{(B)}~=~DASH, \textbf{(C)}~=~regular CSA, \textbf{(D)}~=~a-CSA. 
    Insets show the TPEF along the white, dashed line; the size of the scalebar corresponds to 10\,µm.}
    \label{fig:glia}
\end{figure*}

Figure~\ref{fig:glia}\,(A) shows an example image of a microglia cell in the deep layer, recorded with only the precorrection applied. 
Light scattering in the brain tissue above leads to rather low contrast between structures in Fig.~\ref{fig:glia}\,(A), as illustrated by the plot of signal intensity along the black dashed line, shown in the inset. 
Figures~\ref{fig:glia}\,(B,\,C,\,D) show the same microglia cell after running one out of three AO correction algorithms, respectively, each for 3~iterations of 256~modes.
Figure~\ref{fig:glia}\,(B) shows the cell after application of the correction mask shown the upper left inset, which has been obtained by performing a-CSA with power-overhead in the test mode ($f \sim 2/256$) on top of the precorrection.
The improvement in signal is on the order of a factor of 2~to~3 across the cell body, and processes extending from the cell into its surrounding are starting to become visible, especially when a logarithmic color map is applied (upper right inset).
In Fig.~\ref{fig:glia}\,(C), the cell is shown after performing regular CSA (i.e., without power overhead in the test mode, $f = 1/256$) on top of the precorrection.  The measured modulation SNR was insufficent for algorithm convergence; as is apparent,
the signal quality did not improve compared to the precorrection alone [Fig.~\ref{fig:glia}\,(A)], or even became slightly worse.
Figure~\ref{fig:glia}\,(D) shows the cell after performing DASH (3 iterations of 256 modes, $f = 0.3$) on top of the precorrection.
The signal intensity across the cell body is increased by a factor of about 5; the contrast between structures is clearly enhanced, allowing to distinguish processes extending from the cell body into the surrounding tissue.

Note that for taking the images of Fig.~\ref{fig:glia} we have started from the precorrection of Fig.~\ref{fig:glia}\,(A) simply for illustration purposes, to disentangle the correction of (mild) optical system aberrations from the correction of scattering inside the brain tissue, and for compatibility with the dye-slide simulations and experiments. 
This, however, is not an experimental necessity; typically, images of a similar quality as shown in Fig.~\ref{fig:glia}\,(B,~D) can also be obtained by performing DASH or a-CSA without any precorrection. 

\subsection{Low-SNR scenarios}
\label{sec:lowSNR}

The practicality of every sensorless wavefront correction scheme depends on its robustness with respect to noise. 
It has already been shown that DASH performs well in this regard in comparison to alternative methods~\cite{may2021fast}. 
Let us now compare performances at low-SNR conditions.
The key for all such methods is being able to discern the TPEF signal modulation -- caused by phase-stepping of the test beam -- on top of the noise floor, since this modulation carries the relevant information.
For this, the test mode must contain a significant percentage of the total light power.

In the following, we use numerical simulations to compare which SNR each of the three methods DASH,\footnote{For DASH we slightly increase the power in the  test beam compared to before ($f = 0.3 \rightarrow 0.35$), which has proven helpful in low-SNR situations~\cite{may2021fast}.} a-CSA, and regular CSA\footnote{CSA without controlling the power in individual pixels, as in Ref.~\cite{vellekoop2008phase}.} requires to operate successfully.
To this aim, we simulate shot noise in the PMT readout and then repeatedly execute the methods starting from a decreasing initial signal level. 
We stop repeating once the signal enhancement $\eta$ achieved by a method (after 3 full iterations) drops below a certain threshold $\eta_\text{th}$.
We set this threshold to $\eta_\text{th} = 0.75 \left(\eta_\text{max} - 1\right) + 1$, where $\eta_\text{max}$ is the enhancement for the noise-free case (i.e., simulations as in Section~\ref{sec:systcomp}).
The exact choice of the threshold prefactor (0.75 in our case) is largely arbitrary and uncritical for our conclusions.
We assume the same sample and investigate the same three turbidity scenarios A, B, C as discussed above.

Our simulations show that for weak scattering (A) and correction of $N^2 = 16$ modes, DASH signal enhancement crosses threshold when the photon level (before correction) exceeds around 100 photons.  
a-CSA delivers comparable performance when the test-superpixel contains about 10 times more laser power than any of the other superpixels, corresponding to around 40\% of the total power. 
Regular CSA only crosses threshold for signal levels from around 1000 detected photons onwards, i.e., about 10 times higher than required for DASH.
\newline
For medium turbidity (B) and $N^2 = 64$ modes, DASH crosses threshold at about 100 photons collected per measurement (before correction). 
Comparable performance can be achieved using a-CSA if the test-superpixel amplification factor is around 50, again representing roughly 40\% of the total laser power. 
Regular CSA crosses threshold at around 3\,k photons.
\newline
Finally, for high turbidity (C) and $N^2 = 256$ modes, DASH crosses threshold at around 400 counted photons (before correction). 
Again, comparable performance can be achieved using a-CSA with a test-superpixel amplification factor around 50.
Regular CSA, in contrast, requires about 15\,k photons.

These results are summarized in Table~\ref{tab:table1}.
Of course, in practice the required signal levels will also depend on the nature of the sample, wherefore these numbers can only serve as a rough orientation.
For example, increasing the sample thickness will make it increasingly difficult to obtain a successful correction.

Nevertheless, from our results we can draw three main conclusions.
First, as expected, regular CSA requires a much higher SNR than DASH to operate successfully.
Achieving a higher SNR requires sending more optical power into the sample volume, making regular CSA unfavourable, e.g., for imaging  of fragile specimens.
Second, it is possible to successfully operate a-CSA at the same SNR conditions as DASH, if a sufficient amplification of the test-superpixel can be provided. 
However, it needs to be stressed that if the amplification is realized using a superpixel method (as in Section~\ref{sec:experiment}), discarding light from other pixels, this high performance of a-CSA comes at the price of wasting much optical power. 
For instance, if the hologram features $N{\times}N$ superpixels and the test-superpixel is supposed to contain a fraction $f$ of the total pupil intensity, the incident laser power must be increased by a factor of $(N^2-1) f / (1-f)$ to keep the total optical power in the objective pupil constant. 
Even for a moderate number of modes of 64, i.e., $N = 8$, achieving $f = 0.3$ requires a total power increase by a factor of 27.
This steep scaling with $N$ in practice quickly becomes unfavorable for application of a-CSA in deep tissue imaging, where typically many correctable modes are needed.
Third, for DASH, in contrast, such problems do not arise, since the power in the test mode is independent of $N$ and can be tuned conveniently through the depth of the corresponding grating on the SLM.

\begin{table}[tb]
    \caption{Photon counts (before correction) required at low SNR for comparable performance as in the noise-free case.
    Regular CSA requires much higher signal levels than DASH; a-CSA requires substantial power in the test-superpixel (see main text).}
    \centering
    \begin{tabular}{l c c c}
        \toprule
        \textbf{Turbidity~} & \textbf{low (A)~} & \textbf{medium (B)~} & \textbf{high (C)}\\
        \midrule
        DASH & 100 & 100 & 400\\
        CSA & 1\,k & 3\,k & 15\,k\\
        \bottomrule
         %a-CSA & 40 ($a$ = 80) & 30 ($a$ = 10) & \\
    \end{tabular}
    \label{tab:table1}
\end{table}

\section{Discussion and Summary}

In this work, we have devised a method to compute a 2D phase mask whose effect on a light field is in many ways equivalent to that of a voluminous scatter medium of corresponding scattering mean free path.
This enables software-controlled systematic studies where the degree of scattering is controlled precisely, tunable over a large range, and manual exchange of physical scatter materials is not needed. 

We have used this method to investigate the performance of DASH
for different scales of turbidity, ranging from the ``traditional AO'' regime to the regime of multiply-scattered photons, and compared it to an alternative approach which we call Amplified Continuous Sequential Algorithm (a-CSA). 
a-CSA operates on a square-superpixel mode basis and is shown to work also in low-SNR situations (unlike regular CSA) due to an increased relative power in the test mode.
In practice, this amplification of the test superpixel can be achieved by attenuating all other superpixels, e.g., by adding phase gratings of defined modulation amplitude.
While easy to implement, this approach works only for a limited number of correctable modes due to its inefficient use of light power, and only for a-CSA superpixels consisting of a sufficient number of physical SLM pixels.

We found in both numerical simulations and experiments that for low turbidity DASH initially improves more slowly than a-CSA, pointing to effects of the phase-only light shaping principle behind DASH.
This initial-speed disadvantage of DASH disappears for increasing turbidity.
Furthermore, we observed that independent of the degree of turbidity, from the end of the second iteration onwards, DASH outperforms a-CSA.
We attribute this performance advantage to the fact that DASH can always exploit the full resolution of the SLM, hence minimizes diffraction losses compared to superpixel methods such as a-CSA.
\newline
As an overall tendency, we observed that the signal enhancement achievable using DASH or a-CSA grows for increasing degrees of turbidity, highlighting the potential benefits these algorithms promise for nonlinear imaging in highly scattering environments.

We have illustrated the practical improvements these methods can yield for two-photon microscopy by the example of GFP-microglia 200\,µm deep inside mouse brain tissue.
Furthermore, we emphasize that a-CSA and DASH were performed using identical hardware on the same experimental setup. 
Therefore, depending on the task, it is possible to execute either (or a combination) of both routines on a pure software level. 

A critical challenge for wavefront correction in highly scattering samples remains the size of the corrected field of view (``isoplanatic patch'', IP).
For fixed brain tissue, as shown in Fig.~\ref{fig:glia}, IP diameters are on the order of 20 to 30~µm, whereas live brain tissue tends to scatter photons at larger angles, decreasing the IP size to just a few µm~\cite{may2021simultaneous}.  
Several strategies have been proposed to increase IPs, including multi-conjugate AO~\cite{kam2007modelling, simmonds2013modelling, wu2015numerical} or the application of individual corrections for many sample points~\cite{papadopoulos2020dynamic, may2021simultaneous, blochet2021fast}.
Whether a-CSA and DASH may benefit from such strategies in terms of the IP size is an interesting question for future studies.

%------------------------------------------

\section*{Conflict of Interest Statement}

The authors declare that the research was conducted in the absence of any commercial or financial relationships that could be construed as a potential conflict of interest.

\section*{Author Contributions}

All authors contributed to conceiving the experiment. 
M.S.~conducted the experiments.
M.S.~and A.J.~wrote the manuscript.
A.J.~performed the numerical simulations.
N.B.~contributed to Section~\ref{sec:turbulence} and the Supplementary Materials.

\section*{Funding}

The authors acknowledge funding from the Austrian Science Fund (FWF) under the grants no.~P32146-N36, M~3060-NBL, I3984.

\section*{Acknowledgements}
We would like to thank Kai K.\,Kummer and Jeiny Luna Choconta, Institute for Physiology (Innsbruck Med.~Univ.), for preparing the biological samples. 

% \section*{Supplemental Data}

\section*{Data Availability Statement}

The datasets are available on request.

\bibliography{references.bib}

\clearpage

\appendix
\section*{Supplementary Material}

\subsection{Algorithms for indirect wavefront sensing}

In this Section, we provide the algorithms for DASH and a-CSA in pseudo-code. 
Algorithm~\ref{alg:general} describes general steps that both approaches have in common. 
The function \textsc{wavefront\_sensing} takes as inputs a set of test modes $M$ and a correction pattern $C$, initialized by arrays of ones. 
The algorithm loops $I$ times through the entire set of $N$ modes. 
For each mode $M_n$, the phase stepping procedure \textsc{phase\_step} is executed, which returns an array of complex numbers $U_\text{SLM}$ which carries the amplitude and phase pattern for a particular phase step of the excitation beam.
\textsc{measure\_signal} measures a single TPEF signal $S_p$ for the pattern $U_\text{SLM}$ displayed on the SLM corresponding to a particular phase step $p = 1,\ldots, P$.
From the set of $P$ TPEF measurements per mode the complex-valued coefficient $a_n$ is calculated, containing magnitude and phase estimates of the mode $M_n$. 
Finally, the correction beam $C$ is updated by the function \textsc{update} using the new information ($a_n, M_n$), improving the wavefront correction.

%The sub-function ``measure signal'' represents the physical measurement of the TPEF signal when the pattern $U_{SLM}$ is applied to the SLM. 
DASH and a-CSA differ in how $U_\text{SLM}$ and the corrected beam part $C$ are updated. 
The respective sub-functions are outlined in algorithm~\ref{alg:CSA} for a-CSA and in algorithm~\ref{alg:DASH} for DASH.

The algorithms contain a few variables that are not specifically declared as inputs and can be viewed as ``global'', for instance the number of iterations ($I$), modes ($N$) or phase steps ($P$). A particular variable that needs explanation is the scalar value $f$, which ranges between 0 and 1 and determines the power fraction contained in the test mode. 

Compared to our earlier implementation of DASH~\citep{may2021fast}, the Algorithm~\ref{alg:DASH} outlined below features a minor modification: the normalization of the corrected beam ($|C_{i,n}|$ in Eq.~1 of~Ref.~\cite{may2021fast}) which erases the amplitude information from the corrected field $C_{i,n}$ is replaced by a different scalar normalization factor, which can lead to slightly better performance.

\begin{algorithm}
\caption{General algorithm}\label{alg:general}
\begin{algorithmic}
\State $\mathbf{Inputs{:}}$ $M$, a list of $N$ 2D real-valued input modes of size $N_\text{SLM} \times N_\text{SLM}$; $C$, a 2D array of size $N_\text{SLM} \times N_\text{SLM}$, initialized with ones. \\
\Procedure{wavefront\_sensing}{$M$, $C$}
    \For{$i=1\, \mathbf{to}\, I$}
        \For{$n = 1\, \mathbf{to}\, N$} 
          \For{$p = 1\, \mathbf{to}\, P$} 
                 \State $U_{SLM} \gets \Call{phase\_step}{C, M_n, p}$  
                 \State $S_p \gets \Call{measure\_signal}{U_\text{SLM}}$
                 \EndFor
        \State $a_n \gets \frac{1}{P}\sum^P_{p=1}\sqrt{S_p} ~\exp(-\text i \frac{2\pi}{P}p) $
        \State $C \gets \Call{update}{C, M_n, a_n}$
        \EndFor
    \EndFor      
\State \Return $C$
\EndProcedure
\end{algorithmic}
\end{algorithm}

\begin{algorithm}
\caption{Specific functions for a-CSA}\label{alg:CSA}
\begin{algorithmic}
\Procedure{phase\_step}{$C$, $M_n$, $p$}
    \State $U_{SLM} \gets \sqrt{1-f}~(1 - M_n)~C + \sqrt{f}~M_n~\exp\left(\text i\frac{2\pi}{P}p\right)$  %C^*$ 
\State \Return $U_\text{SLM}$
\EndProcedure
\\
\Procedure{update}{$C$, $M_n$, $a_n$}
    \State $C \gets C~(1 - M_n) + M_n~\exp{\left(-\text i~M_n~\mathrm{angle}(a_n)\right)}$   
    \State \Return $C$
\EndProcedure
\end{algorithmic}
\end{algorithm}

\begin{algorithm}
\caption{Specific functions for DASH}\label{alg:DASH}
\begin{algorithmic}
\Procedure{phase\_step}{$C$, $M_n$, $p$}
    \State $U_\text{SLM} \gets \sqrt{1-f}~C + \sqrt{f} \exp\left(\text i(M_n + \frac{2\pi}{P}p)\right)$
    \State \Return $\exp{\left(\text i~\mathrm{angle}(U_\text{SLM})\right)}$
\EndProcedure
\\
\Procedure{update}{$C$, $M_n$, $a_n$}
    \State $C \gets C + a_n^*~\exp{\left(\text i~ M_n\right)}$   
    %\Comment Updating the correction pattern
    \State $\gamma \gets \sqrt{\frac{1}{N_\text{SLM}^2}\sum|C|^2}$
    %\Comment Calculate normalization
    \State $C \gets \frac{C}{\gamma}$
    \Comment Normalization
    
\State \Return $C$
\EndProcedure
\end{algorithmic}
\end{algorithm}

\subsection{Construction of scatter mask in the numerical simulation and experiment}

Algorithm~\ref{alg:scat} describes the procedure of constructing the scattering phase mask displayed at the SLM.
We denote the RMS value of a scattering phase mask by $a_\text{scat}$. 
If the phase values of the mask are normal-distributed or, for any distribution, if $a_\text{scat}$ is sufficiently small~\cite{wang2011scattering}, the relation between the scatterer thickness and $a_\text{scat}$ is simply $L/l_s = \sqrt{a_\text{scat}}$.

The function \textsc{make\_scatterer} takes the following inputs: the desired spatial frequency content characterized by $\sigma$, the side length of the phase mask $N_\text{scat}$, and the desired RMS phase magnitude of the scatterer $a_\text{scat}$. 
The pixel indices are given by $x$ and $y$. 
The function \textsc{rand} creates equally distributed random numbers in the interval $[0,1[$.
\textsc{fft} represents the fast Fourier transform and \textsc{norm} normalizes a real-valued array to an RMS value of $a_\text{scat}$.

\begin{algorithm}
\caption{Scatterer calculation}\label{alg:scat}
\begin{algorithmic}
\Procedure{make\_scatterer}{$\sigma$, $N_\text{scat}$, $a_\text{scat}$}
    \State $W \gets \exp{\left(-(x^2 + y^2)/(2\sigma^2)\right)}$  %C^*$ 
    %\Comment{constructing Gaussian weighting}
    \State $A \gets \exp\left(\text i~2\pi~\Call{rand}{N_\text{scat}, N_\text{scat}}\right) \cdot W$
    %\Comment{initialize field of random phase values}
    \State $A[1] \gets 0 $ 
    \Comment{discard DC term}
    \State $\Phi_\text{scat} \gets \mathrm{real}\left(\Call{fft}{A}\right)$
    %\Comment{take real part of Fourier transform}
    \State $\Phi_\text{scat} \gets \Call{norm}{\Phi_\text{scat}, a_\text{scat}}$
    %\Comment{normalization to an RMS value of $a_\text{scat}$}
\State \Return $\Phi_\text{scat}$
\EndProcedure
\end{algorithmic}
\end{algorithm}

\subsection{Animations of DASH and CSA correction procedures}
\begin{figure}
    \centering
    \includegraphics[width = \columnwidth]{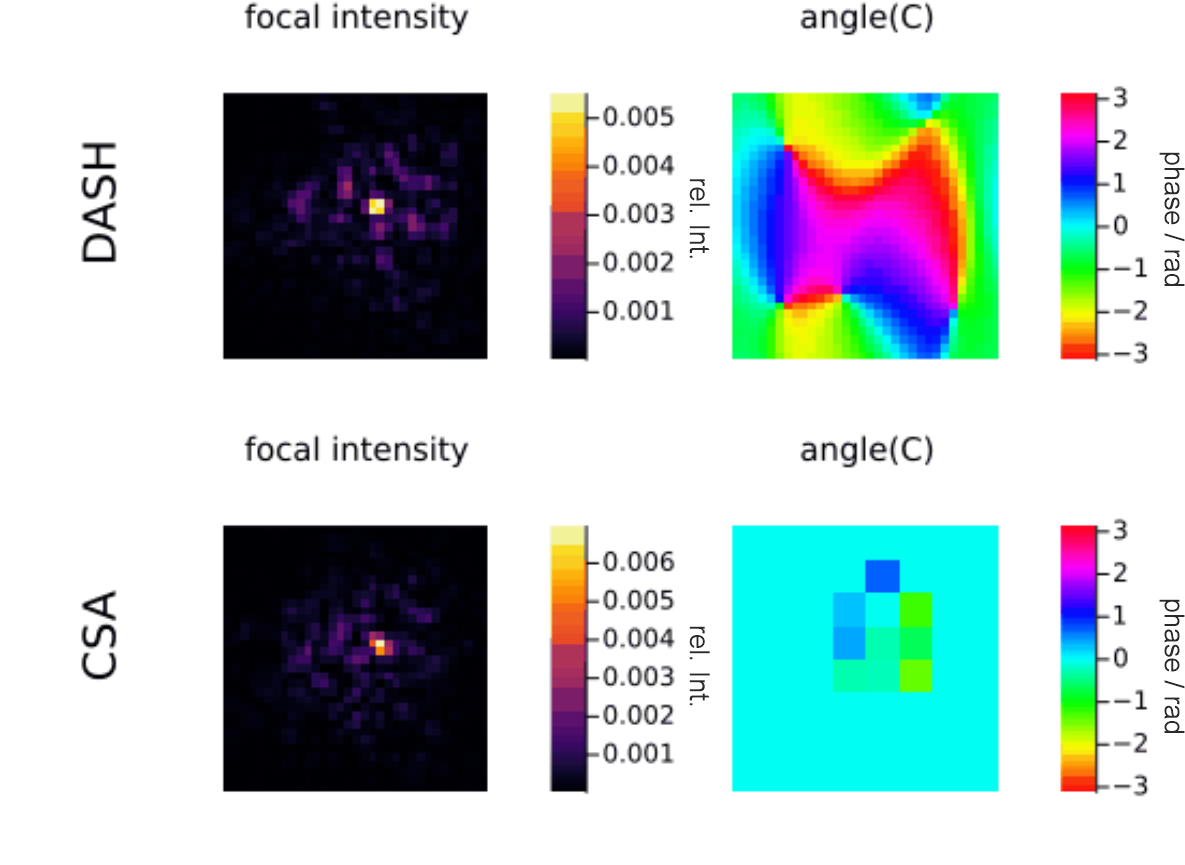}
    \caption{\textbf{Exemplary frame from the GIF animation for scenario \textbf{B}.} Focus quality and phase correction pattern after optimization of the 10th mode in the first optimization iteration.}
    \label{fig:animation}
\end{figure}

We provide three GIF animations visualizing the progress of correction for DASH and CSA\footnote{Note that for the noise-free case as discussed here (SNR${}\rightarrow\infty$) the performance of regular and a-CSA is identical, wherefore we use the more general label ``CSA''.} 
for the three scenarios A, B and C as discussed in the main document.
The animations show the evolution of focus quality (left column) and corresponding correction phase masks over three measurement iterations. The scalebar for the intensity images is normalized to the maximum intensity of an aberration-free focus. 
An exemplary frame of the animation for scenario B is shown in Fig.~\ref{fig:animation}.

\end{document}